\documentclass[
aps,
prc,
superscriptaddress,
floatfix,
nofootinbib,
twocolumn
]{revtex4-2}

\usepackage{amsmath,amssymb,amsfonts,physics,graphicx,float}

\usepackage[setpagesize=false]{hyperref}
\allowdisplaybreaks[4]

\usepackage[labelformat=simple]{subcaption} 

\usepackage{ulem}
\usepackage{color}
\definecolor{ar}{rgb}{1.0, 0.01, 0.24}
\definecolor{al}{rgb}{0.82, 0.1, 0.26}
\definecolor{ev}{rgb}{0.56, 0.0, 1.0}

\begin{document}

\title{
Constraints on the strength of first-order phase transition and its relation to nucleon mass
}

\author{Bikai Gao}
\email{bikai@rcnp.osaka-u.ac.jp}
\affiliation{Research Center for Nuclear Physics, The University of Osaka, Ibaraki, Osaka 567-0047, Japan}

\date{\today}

\begin{abstract}
We investigate constraints on the strength of first-order phase transitions in neutron star matter and its relation to the origin of nucleon mass. By combining the parity doublet model for the hadronic phase, the Nambu-Jona-Lasinio model for quark matter, and the integral constraint framework for intermediate densities, we construct  equation of state spanning the full density range relevant to neutron stars. Our approach systematically explores how the chiral invariant mass $m_0$ affects the allowable properties of first-order quark-hadron phase transitions. Through comparison with recent neutron star observations, we establish a inverse correlation between the allowed phase transition strength and the chiral invariant mass. Our results demonstrate a direct connection between fundamental questions about the microscopic origin of nucleon mass and macroscopic neutron star observables, providing a novel astrophysical probe of chiral dynamics and QCD physics under extreme conditions. 
\end{abstract}
\maketitle

\section{INTRODUCTION}

The equation of state (EOS) of dense nuclear matter is crucial for understanding neutron star (NS) structure and evolution. Of particular significance is the potential existence of phase transitions (PT) within NS interiors, where hadronic matter may transit to exotic states such as quark matter through crossover or 1st-order PT. First-order PTs are characterized by discontinuous jumps in energy density and baryon number density while maintaining continuity in pressure and chemical potential. Unlike continuous crossovers (\cite{Baym:2017whm, Baym:2019iky, Kojo:2021wax}),  the discontinuities introduced by 1st-order PTs can significantly alter NS observation properties, producing distinctive signatures in both electromagnetic and gravitational wave observations \cite{Lenzi:2012xz, Benic:2014jia, Christian:2023hez, Gao:2024lzu,Fujimoto:2022xhv,Huang:2022mqp,Guo:2023som}. In cases of sufficiently strong 1st-orderPTs, the mass-radius ($M$-$R$) relationship may even develop disconnected branches, potentially leading to ``twin stars''---NSs with similar masses but significantly different radii \cite{Kampfer:1981yr, Glendenning:1998ag,  Benic:2014jia, Alford:2017qgh, Most:2018hfd}.

However, the strength of first-order phase transitions is not an arbitrary parameter but is intimately connected to the fundamental properties of nuclear matter, such as the nucleon mass. The  nucleon mass is intimately connected with the phenomenon of chiral symmetry breaking. In the limit where quark masses vanish, QCD possesses an approximate chiral symmetry that is spontaneously broken in the vacuum, generating the bulk of hadron masses through the formation of quark condensates. However, this picture becomes increasingly complex under the extreme conditions found in NS cores, where densities can reach 5-10 $n_0$  ($n_0$: nuclear saturation density) and the restoration of chiral symmetry may occur. Understanding how this restoration proceeds, and what fraction of nucleon mass persists even when chiral symmetry is fully restored—characterized by the chiral invariant mass \cite{Weinberg:1969hw, Detar:1988kn, Jido:2001nt}—represents a fundamental challenge in our understanding of strong interaction physics.

The parity doublet model (PDM) \cite{Detar:1988kn, Jido:2001nt, Minamikawa:2020jfj, Zschiesche:2006zj,PhysRevC.77.025803,PhysRevC.82.035204,Motohiro:2015taa,Marczenko:2020jma,Marczenko:2022hyt, Kong:2023nue,Minamikawa:2023ypn, Gao:2024chh, Gao:2024lzu,Gao:2024mew} provides a particularly useful framework for addressing these questions. In this approach, nucleons and their opposite-parity partners are treated as chiral partners whose masses become degenerate when chiral symmetry is restored at high densities. The key parameter in this model is the chiral invariant mass $m_0$, which represents the portion of the nucleon mass that persists even when chiral symmetry is fully restored. This parameter influences the stiffness of the EOS. By varying $m_0$, one can systematically explore how different assumptions about the origin of nucleon mass affect the nuclear EOS and, consequently, the allowable properties of first-order phase transitions by confronting with recent NS observations\citep{LIGOScientific:2017vwq,LIGOScientific:2017ync,LIGOScientific:2018cki,Miller:2019cac,Miller:2021qha}.

The theoretical description of matter across the wide density range relevant to NSs presents significant challenges due to the different physical regimes that emerge as density increases. At low densities near nuclear saturation, effective hadronic theories provide reliable descriptions of nuclear matter properties, with nucleons serving as the fundamental degrees of freedom. However, as density increases beyond 2-3$n_0$, the validity of purely hadronic descriptions becomes questionable\cite{Baldo:1997ag,Li:2019xxz,Drischler:2021kxf}, and quark degrees of freedom become increasingly relevant as the internal structure of hadrons begins to overlap and merge \cite{Baym:2017whm,Baym:2019iky,Kojo:2021wax}. At densities exceeding approximately 5$n_0$, quarks are expected to be deconfined, and the system transitions to a quark matter phase. In this regime, effective quark models such as the Nambu-Jona-Lasinio (NJL) model~\cite{Hatsuda:1994pi,Baym:2017whm, Baym:2019iky, Kojo:2021wax,Gao:2022klm, Yuan:2023dxl, Gao:2024chh, Gao:2024lzu,Gholami:2024ety} provide valuable tools for describing the properties of quark matter. These models capture important features of QCD including dynamical chiral symmetry breaking and color superconductivity at high densities. At extremely high densities where the baryon number density reaches $n_B \geq 40n_0$, asymptotic freedom suggests that perturbative QCD should eventually become applicable, providing a first-principles description of strongly interacting matter \cite{Kurkela:2009gj,Gorda:2021znl}. However, the densities achievable in NS cores are unlikely to reach this perturbative regime.

In the intermediate density region roughly between 2$n_0$ and 5$n_0$ still remains challenging. In this regime, neither purely hadronic nor purely quark descriptions may be entirely adequate, and the transition between these different phases of matter likely occurs. It is precisely in this poorly understood density range where first-order phase transitions are most likely to manifest, making reliable theoretical treatment of this region crucial for understanding NS structure and the constraints that observations can place on fundamental physics. Recent theoretical developments have provided new tools to address these challenges through model-independent approaches~\cite{Komoltsev:2021jzg} based on fundamental principles of thermodynamic stability and causality. This method can constrain the allowable EOS in the intermediate density regime without requiring detailed microscopic knowledge.

In this work, we then combine the PDM for the hadronic phase, the NJL-type quark model for quark matter, and the integral constraint framework for the intermediate density region to construct EOS that span the full density range relevant to NSs. By systematically varying the chiral invariant mass $m_0$ and the strength of first-order transitions, we determine the parameter combinations that remain consistent with current observational constraints. Our analysis reveals how fundamental questions about the origin of nucleon mass translate into observable NS properties, providing a novel astrophysical probe of chiral dynamics and QCD physics under extreme conditions.

This paper is organized as follows. In Section~\ref{sec:pdm}, we describe the hadronic equation of state based on the parity doublet model. Section~\ref{sec:integral_constraint} introduces the integral constraint framework, which provides model-independent bounds on the EOS. In Section~\ref{sec:construction}, we present our methodology for constructing first-order phase transitions by combining the three theoretical frameworks and systematically varying the transition strength while maintaining thermodynamic consistency. We analyze the resulting mass-radius relations and compare them with current neutron star observations to establish constraints on the allowable phase transition parameters. Section~\ref{sec:summary} summarizes our main findings and discusses their implications for understanding the origin of nucleon mass and QCD physics under extreme conditions. The technical details of the quark matter EOSbased on the Nambu-Jona-Lasinio model are provided in Appendix~\ref{app:quark}.

\section{Hadronic EOS based on Parity doublet model} \label{sec:pdm}
We employ the parity doublet framework for hadronic EOS as discussed in (\citep{Gao:2022klm, Minamikawa:2023eky}). In this approach, two types of nucleons with positive and negative parity states are treated as chiral partners, and their masses become degenerate when chiral symmetry is restored at high densities. This mass, known as the chiral invariant mass, is a key parameter in these models, significantly influencing the stiffness of the EOS. In particular, a larger $m_0$ results in weaker $\sigma$ couplings to nucleons since the mass of nucleon is not completely derived from the $\sigma$ fields. Correspondingly, the couplings to $\omega$ fields are reduced since at the nuclear saturation density $n_0$, the repulsive contributions of the $\omega$ fields must be counterbalanced by the attractive $\sigma$ contributions. Beyond densities greater than $n_0$, the $\sigma$ fields decrease while the $\omega$ fields increase, leading to an imbalance. As a consequence, a larger $m_0$ weakens the $\omega$ fields and softens EOS at supranuclear densities. Typical PDMs are $\sigma$-$\omega$ type mean field models, with some works also incorporating the isovector scalar meson $a_0(980)$, which is believed to appear in asymmetric matter, such as in neutron stars. However, as investigated in Ref.~\cite{Kong:2023nue}, the inclusion of the $a_0(980)$ has a negligible impact on the properties of neutron stars, resulting in only a slight increase in the radius of less than $1 \rm km$. In this study, we then consider the PDM model with $N_f=2$ and include the vector meson mixing, such as the $\omega^2\rho^2$ interaction, as described in Ref.~\cite{Gao:2022klm}. 

The thermodynamic potential of the model in the mean-field approximation is calculated as
\begin{equation}
\begin{aligned}
\Omega_{\mathrm{PDM}}&=V_\sigma-V\left(f_{\pi}\right)\\
&+V_\omega+V_\rho + V_{\omega\rho} +\sum_{i=+,-} \sum_{x=p, n} \Omega_{x} \ ,\label{Eq:Omega PDM}
\end{aligned}
\end{equation}
where $i = +, -$ denote the positive-parity ordinary nucleon $N(939)$ and negative-parity excited nucleon $N^{*}(1535)$. The mean-field potential $V(\sigma)$, $V_\omega$, $V_\rho$ and $V_{\omega \rho}$ are given by
\begin{equation}
\begin{aligned}
V(\sigma) = -\frac{1}{2}\bar{\mu}^{2}\sigma^{2} &+ \frac{1}{4}\lambda_4 \sigma^4 -\frac{1}{6}\lambda_6\sigma^6 - m_{\pi}^{2} f_{\pi}\sigma\ \ , \\
V_\omega&=-\frac{m_\omega^2}{2} \omega^2 \ ,\\
V_\rho&=-\frac{m_\rho^2}{2} \rho^2 \\
V_{\omega \rho}&=-\lambda_{\omega\rho}(g_{\omega NN}\omega)^2(g_{\rho NN}\rho)^2\ ,\label{Eq:PDM potential}
\end{aligned}
\end{equation}
with $f_{\pi}$ the pion decay constant. Here, $\bar{\mu}, \lambda_{4}, \lambda_{6}$ and $\lambda_{\omega\rho}$ are parameters to be determined and the kinetic part of the thermodynamic potential $\Omega_x$ reads
\begin{equation}
\begin{aligned}
\Omega_x= -2 \int^{k_x^{\pm}} \frac{\mathrm{d}^3 \mathbf{p}}{(2 \pi)^3}\left(\mu_x^*-E_{\mathbf{p}}^i\right),\ \label{Eq: PDM kinetic part} 
\end{aligned}
\end{equation}
with $E_{{\bf p}}^{i} = \sqrt{{\bf p}^{2} + m_{\pm}^{2}}$ is the energy of relevant nucleon with mass $m_{\pm}$ and momentum ${\bf p}$, and $k_{x}^{\pm}=\sqrt{(\mu^{*}_{x})^2-m_{\pm}^2}$ is the fermi momentum for the relevant particle, in which $\mu_x^{*}$ is the effective chemical potential. We notice that we use the no-sea approximation, assuming that the structure of the Dirac sea remains the same for the vacuum and medium. 

The masses of the positive- and negative-parity chiral partners are given by
\begin{equation}
\begin{aligned}
m_{ \pm}=\frac{1}{2}\left[\sqrt{\left(g_1+g_2\right)^2 \sigma^2+4\left(m_0\right)^2} \mp\left(g_1 - g_2\right) \sigma\right]\ ,\label{Eq: PDM mass}
\end{aligned}
\end{equation}
where $\pm$ sign denotes parity. The spontaneous chiral symmetry breaking yields the mass splitting between the two baryonic parity partners in each parity doublet. When the symmetry is restored, the masses in each parity doublet become degenerate: $m_{\pm}(\sigma=0)=m_0$. The positive-parity nucleons are identified as the positively charged and neutral $N(939)$ states: proton $(p)$ and neutron $(n)$. Their negative-parity counterparts, denoted as $p^{*}$ and $n^{*}$, are identified as $N(1535)$ resonance. For a given chirally invariant mass, $\mathrm{m}_0$, the parameters $\mathrm{g}_1$ and $\mathrm{g}_2$ are determined by the corresponding vacuum masses, $m_N=939 \mathrm{MeV},  m_{N^*}=1535 \mathrm{MeV}$. The effective chemical potentials for nucleons and their chiral partners are given by
\begin{equation}
\begin{aligned}
\mu_p=\mu_p^{*}&=\mu_Q+\mu_B-g_{\omega NN} \omega- \frac{1}{2}g_{\rho NN} \rho \ , \\
\mu_n=\mu_n^{*}&=\mu_B - g_{\omega NN} \omega+ \frac{1}{2}g_{\rho NN} \rho \ .\label{Eq: PDM chemicalPoten}
\end{aligned}
\end{equation}
 
\begin{figure}[htp]
\centering
\includegraphics[width=1\hsize]{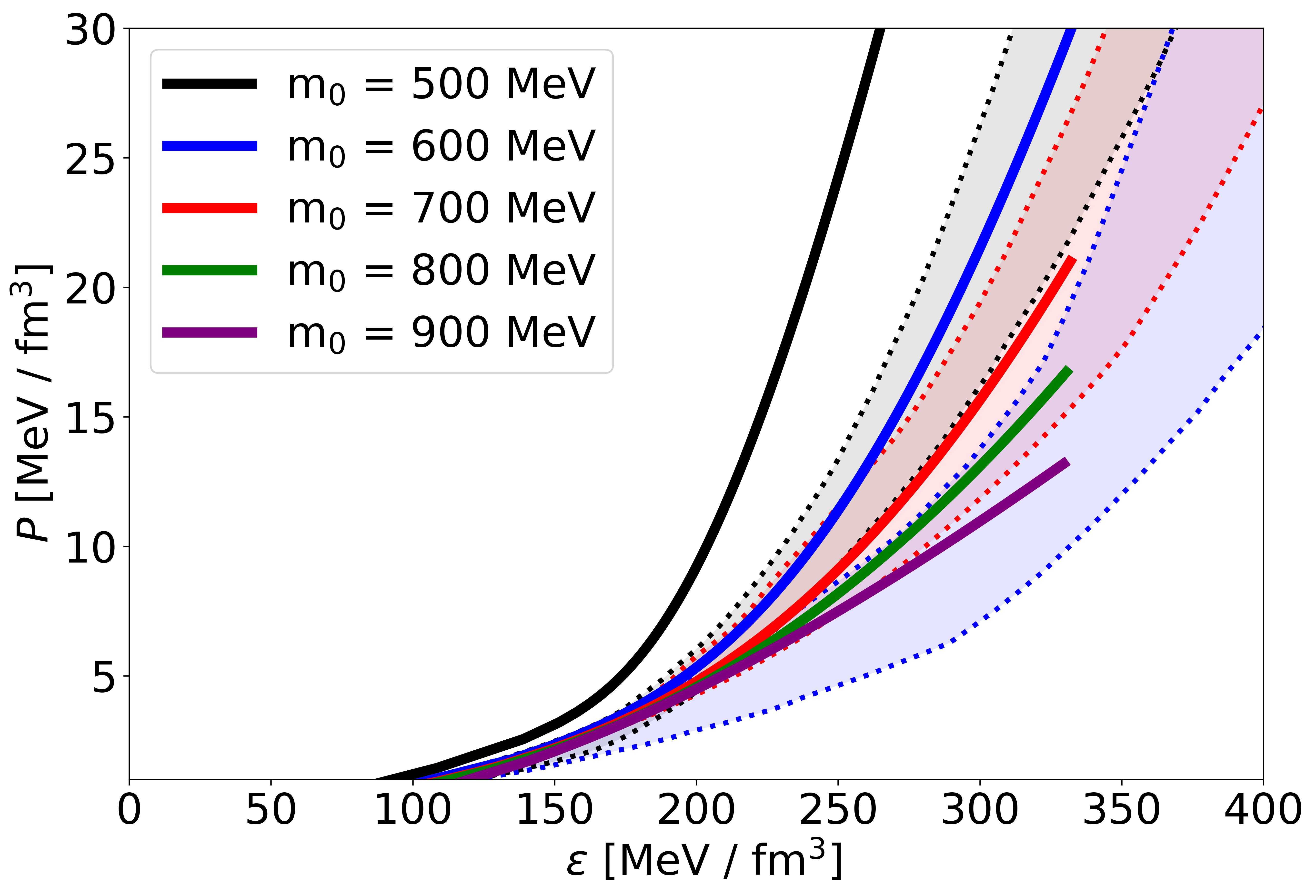}
\caption{EOS in the PDM for different chiral invariant mass $m_0$ with fixed slope parameter $L_0 = 57.7$ MeV. Red band show  data-driven EOSs inferred from deep learning (edged by dotted red curves \cite{Fujimoto:2021zas}) and Bayesian (edged by blue dotted \cite{Raaijmakers:2021uju} and black dotted curves \cite{Ozel:2015fia,Bogdanov:2016nle}) analyses.}
\label{fig:EOS_compare}
\end{figure}

The total thermodynamic potential of the hadronic matter in neutron stars is obtained by including the effects of leptons as
\begin{equation}
\begin{aligned}
\Omega_{\mathrm{H}}=\Omega_{\mathrm{PDM}} + \Omega_{e},
\end{aligned}
\end{equation}
where $\Omega_{e}$ is the thermodynamic potentials for electrons given by 
\begin{equation}
\begin{aligned}
\Omega_{e}=-2 \int^{k_F} \frac{\mathrm{d}^3 \mathbf{p}}{(2 \pi)^3}\left(\mu_l-E_{\mathbf{p}}^l\right),
\end{aligned}
\end{equation}
Finally, we have the pressure in hadronic matter as
\begin{equation}
P_{\mathrm{H}}=-\Omega_{\mathrm{H}}.
\end{equation}
Using the explicit parameter sets determined in Ref.~\cite{Minamikawa:2020jfj}, with fitting to the pion decay constant $f_\pi=93$ MeV and hadron masses, as well as to the normal nuclear matter properties (Saturation density $n_0=0.16$ fm$^{-3}$ , binding energy $B_0=16$ MeV, incompressibility $K_0 = 240$ MeV, and symmetry energy $S_0 = 31$ MeV), we can calculate the corresponding EOS in the hadronic phase for different choices of the chiral invariant mass $m_0$ as shown in Fig.~\ref{fig:EOS_compare}. The resulting EOS can be compared with the data-driven EOSs inferred from deep learning (edged by dotted red curves \cite{Fujimoto:2021zas}) and Bayesian (edged by blue dotted \cite{Raaijmakers:2021uju} and black dotted curves \cite{Ozel:2015fia,Bogdanov:2016nle}) analyses. For $m_0=500$ MeV, we find that the EOS becomes excessively stiff, falling outside the bounds of all data-driven analyses. Therefore, we restrict our investigation to the  $m_0$ range larger than 600 MeV for this study. The resulting EOS from our PDM provides the low-density anchor for our unified NS EOS, which will be connected to higher densities as described in subsequent sections.

\section{Integral constraint}\label{sec:integral_constraint}

Our construction of 1st-orderphase transitions employs the integral constraint framework introduced in Ref.~\cite{Komoltsev:2021jzg}. This model-independent approach requires matching conditions between low-density and high-density boundary points in the EOS. We define the low-density boundary at ($P_L, n_L, \mu_L$) and the high-density boundary at ($P_H, n_H, \mu_H$), where $P$ denotes pressure, $n$ is baryon number density, and $\mu$ represents the baryon chemical potential. For demonstration purposes, we use specific boundary values: the low-density point is determined from our PDM calculation at $n_L = 1.5n_0$ with $m_0 = 600$ MeV, while the high-density boundary is taken from perturbative QCD (PQCD) calculations with $\mu_H = 2600$ MeV, $n_H = 6.47$ fm$^{-3}$, and $P_H = 3823$ MeV/fm$^3$.

\begin{figure}[htp]
\centering
\includegraphics[width=1\hsize]{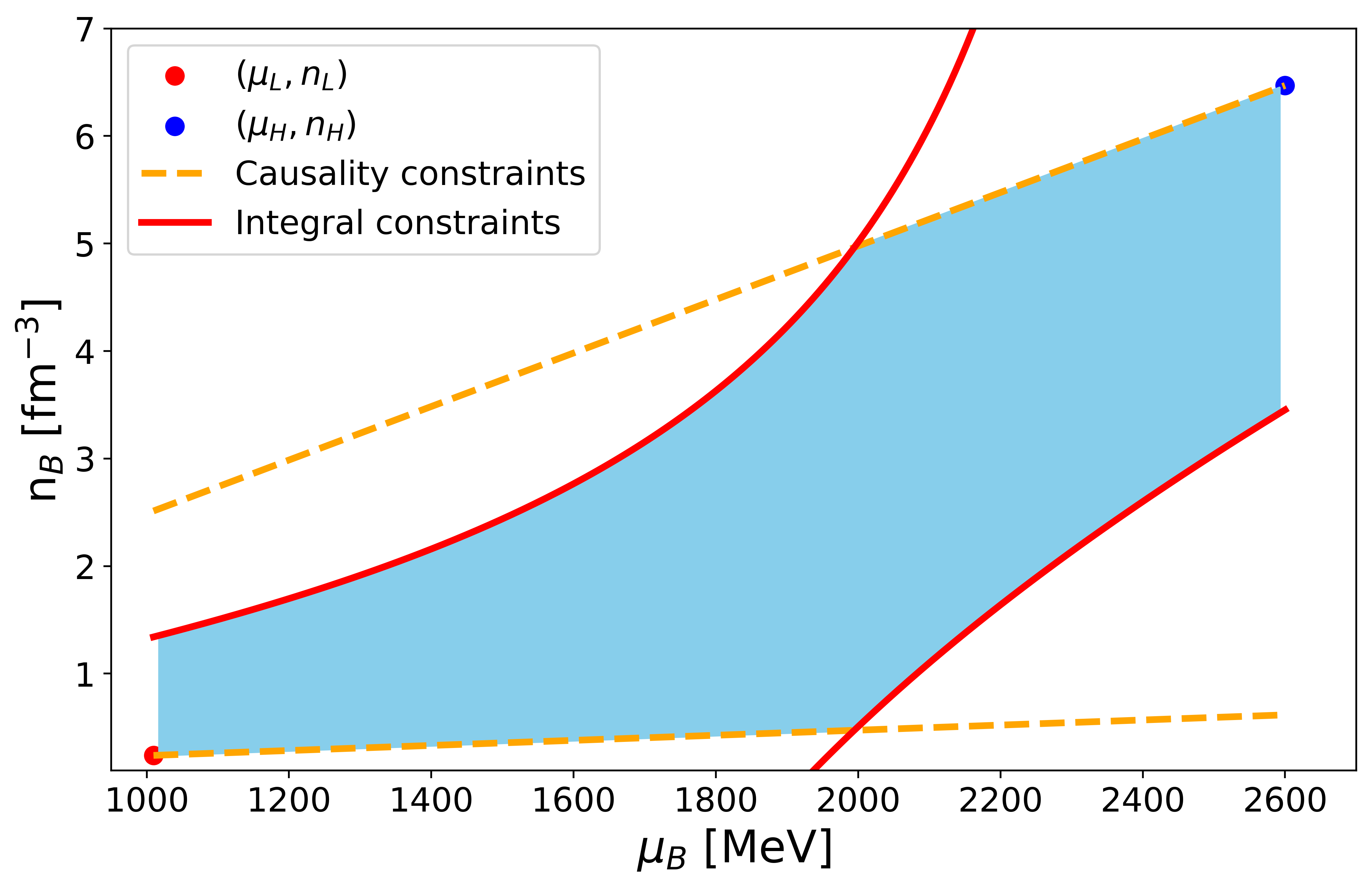}
\caption{Constrained region in the $\mu_B$-$n_B$ plane. The orange curves represent causality boundaries ($c_s^2 = 1$) begin from the low-density point $(\mu_L, n_L)$ and high-density point $(\mu_H, n_H)$. The red curves show integral constraint boundaries where $\Delta P_{\rm min} = \Delta P$ and $\Delta P_{\rm max} = \Delta P$.  }
\label{fig:n_muB}
\end{figure}

The physically allowable region between these boundaries is constrained by two fundamental requirements. First, thermodynamic stability demands that pressure increases monotonically with chemical potential. Second, causality requires that the sound velocity satisfies $c_s^2 \leq 1$. For any interpolating function $n_B(\mu)$ connecting the boundary conditions, the causality constraint
\begin{align}
c_s^{-2} = \frac{\mu}{n}\frac{\partial n}{\partial \mu} \geq 1
\end{align}
determines the minimum slope of $n_B(\mu)$. By solving the limiting case $c_s^2 = 1$, we obtain two causality boundaries begin from $(\mu_L, n_L)$ and $(\mu_H, n_H)$ in the $n_B$-$\mu_B$ plane, shown as orange curves in Fig.~\ref{fig:n_muB}.

Additionally, thermodynamic consistency requires
\begin{align}
\int^{\mu_H}_{\mu_L} n_B(\mu) d\mu = P_H - P_L = \Delta P.
\end{align}
For each point in the $n_B$-$\mu_B$ plane, we evaluate the absolute minimum $\Delta P_{\rm min}$ and maximum $\Delta P_{\rm max}$ achievable if the EOS passes through that point. The detailed construction of these  areas is described in Ref.~\cite{Komoltsev:2021jzg}. Points where $\Delta P_{\rm min} > \Delta P$ or $\Delta P_{\rm max} < \Delta P$ violate the thermodynamic constraint and are excluded, defining additional boundaries shown as red curves in Fig.~\ref{fig:n_muB}. The intersection of all constraints yields the physically allowed region, shown as the blue shaded area in Fig.~\ref{fig:n_muB}. For each given baryon chemical potential, the maximum baryon number density is 
\begin{align}
&n_{\max }(\mu_B)\nonumber \\
&= \begin{cases}\frac{\mu_B^3 n_L-\mu_B \mu_L\left(\mu_L n_L+2 \Delta p\right)}{\left(\mu_B^2-\mu_H^2\right)\mu_L}, & \mu_L \leq \mu_B <\mu_c, \\ n_H \mu_B / \mu_H, & \mu_c \leq \mu_B \leq \mu_H,\end{cases}
\end{align}
and the minimum allowed density is 
\begin{align}
&n_{\min }(\mu_B) \nonumber\\
&= \begin{cases}n_L \mu_B / \mu_L, & \mu_L \leq \mu_B \leq \mu_c, \\ \frac{\mu_B^3 n_H-\mu_B \mu_H\left(\mu_H n_H-2 \Delta p\right)}{\left(\mu_B^2-\mu_L^2\right) \mu_H}, & \mu_c<\mu_B \leq \mu_H .\end{cases}
\end{align}
with $\mu_c$ is the intersection point of the causality boundary and the integral constraint boundary.

\begin{figure}[htp]
\centering
\includegraphics[width=1\hsize]{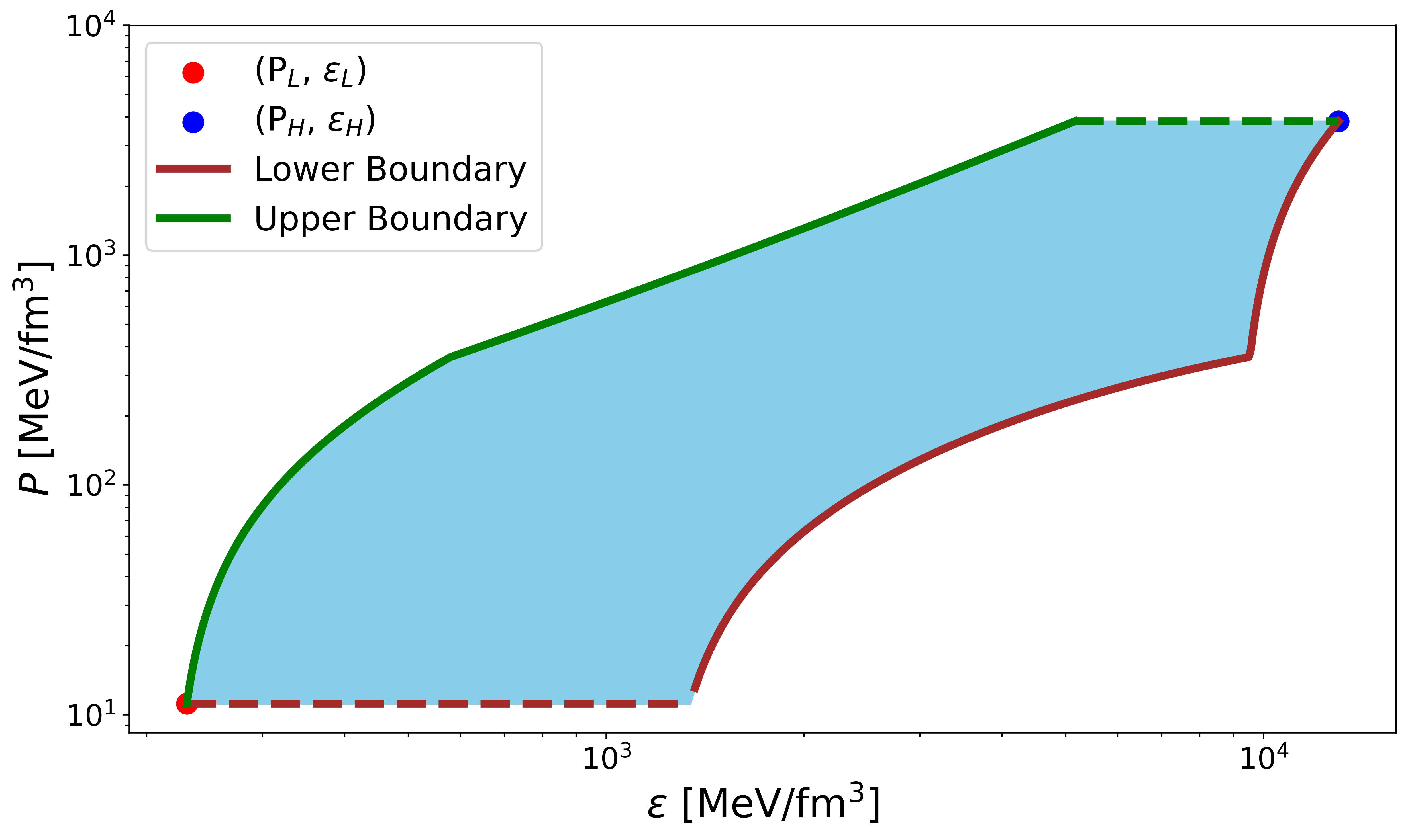}
\caption{Constrained region in the pressure-energy density ($P$-$\varepsilon$) plane, obtained by mapping the allowed parameter space from Fig.~\ref{fig:n_muB}. The upper boundary (green curve) represents the stiffest possible EOS connection between the low-density and high-density points, while the lower boundary (brown curve) corresponds to the softest possible connection. The blue shaded region encompasses all physically allowable EOS that satisfy causality and thermodynamic constraints.}
\label{fig:p_e_constrained}
\end{figure}

By integrating over this constrained region, we can map the allowed parameter space from the $n_B$-$\mu_B$ plane to the pressure-energy density ($P$-$\varepsilon$) plane, as illustrated in Fig.~\ref{fig:p_e_constrained}. The constrained region in the $P$-$\varepsilon$ plane exhibits several important characteristics. The upper boundary (green curves) represents the stiffest possible connection between the low-density and high-density points, yielding the maximum pressure for a given energy density. Conversely, the lower boundary (brown curve) corresponds to the softest possible connection, providing the minimum pressure at each energy density. Both the upper and lower boundaries consist of three distinct segments. The dashed portions originate from density discontinuities that begin at the low-density and high-density boundary points, as evident from the vertical jumps in Fig.~\ref{fig:n_muB}. The solid curves arise from the combined effects of the causality condition and integral constraint. The kink appearing in the middle of each solid curve marks the intersection point where the thermodynamic constraint boundaries (red curves) meet the causality boundaries (orange curves) in the $n_B$-$\mu_B$ plane shown in Fig.~\ref{fig:n_muB}.

This mapping provides a  framework for understanding the physically allowable EOS connections between low density and  high density phases, with the upper boundary representing the stiffest possible EOS and the lower boundary the softest possible EOS that satisfy all fundamental constraints.  This approach offers several  advantages over traditional approaches. First, it remains model-independent, relying only on fundamental physical principles; Second, it provides  constraints without requiring detailed knowledge of microscopic physics; and also it naturally accommodates various scenarios of quark-hadron transitions without assuming their specific nature, whether smooth crossover or 1st-orderPT.


\section{Construction of the 1st-order phase transition}\label{sec:construction}

In this section, we analyze the relation between the allowable strength of 1st-order PTs and the chiral invariant mass $m_0$, using constraints from recent NS observations.

To characterize the high-density regime relevant to NS cores, we employ the NJL-type quark model \cite{Hatsuda:1994pi,Baym:2017whm, Baym:2019iky, Kojo:2021wax, Yuan:2023dxl, Gao:2024chh, Gao:2024lzu,Gholami:2024ety,Gao:2024jlp,Yuan:2025dft} to obtain the boundary conditions $(P_H, n_H, \mu_H)$ and the corresponding quark matter EOS above density $n_H$. At NS core densities, quarks become the fundamental degrees of freedom. However, with baryon chemical potentials $\mu_B \sim 1.5$-$2$ GeV, the system enters a strongly coupled regime where $\alpha_s \sim \mathcal{O}(1)$, rendering perturbative QCD unreliable.

The NJL model provides an effective description that captures important QCD features—dynamical chiral symmetry breaking and color superconductivity—through four-fermion interactions, while treating gluonic degrees of freedom implicitly. Our implementation includes three quark flavors (up, down, strange) and three colors, with two key parameters governing the dense matter behavior: the vector coupling $g_V$, which controls repulsive vector-channel interactions, and the diquark coupling $H$, which determines the strength of attractive correlations leading to color superconductivity. Following \cite{Baym:2017whm, Kojo:2021wax}, we treat $g_V$ and $H$ as independent parameters whose allowed ranges are constrained by fundamental physical requirements and astronomical observations of massive NSs.

\subsection{Methodology}

We begin by establishing a systematic procedure for connecting hadronic and quark matter EOS. For a fixed hadronic EOS in the low-density region and a fixed quark EOS in the high-density region, we determine whether these can be connected while satisfying fundamental constraints of thermodynamic stability and causality as introduced in Sec.~\ref{sec:integral_constraint}. When such a connection is possible, we obtain both upper and lower boundaries in the intermediate density region between $n_L$ and $n_H$, with the upper boundary representing the stiffest physically allowable connection between the low and high-density regimes similar to Fig.~\ref{fig:p_e_constrained}.

\begin{figure}[htp]
\centering
\includegraphics[width=1\hsize]{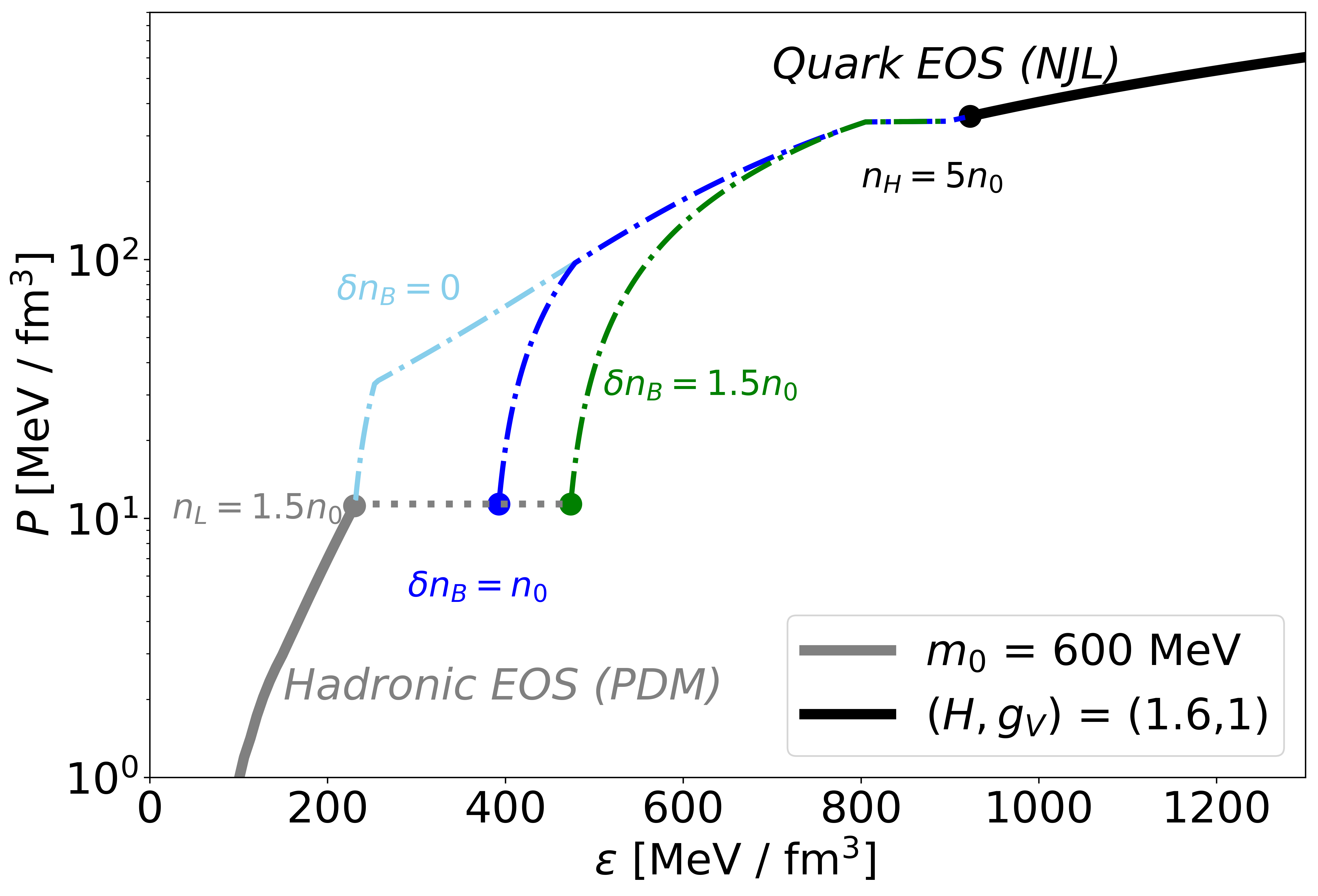}
\caption{EOS connections between the hadronic phase described by the PDM with $m_0=600$ MeV (thick grey curve) and the quark phase with $(H, g_V)/G = (1.6, 1)$ (thick black curve) in the pressure versus energy density plane. The colored curves represent the upper boundaries (stiffest possible EOS) for different strengths of 1st-orderPTs: $\delta n_B = 0$ (skyblue curve), $\delta n_B = n_0$ (blue curve), and $\delta n_B = 1.5n_0$ (green curve). }
\label{fig:p_e}
\end{figure}

As illustrated in Fig.~\ref{fig:p_e}, we employ a hadronic EOS with $m_0 = 600$ MeV (thick grey curve) and a quark EOS with $(H, g_V)/G = (1.6, 1)$ (thick black curve). For demonstration purposes, we set the initial matching density $n_L=1.5 n_0, n_H = 5 n_0$ and determine the corresponding pressure $P_L, P_H$, chemical potential $\mu_L, \mu_H$ from the PDM and the NJL-type quark model, respectively.  This yields the upper boundary shown by the thin skyblue curve in Fig.~\ref{fig:p_e}. We then model 1st-order PTs by introducing a density jump while preserving thermodynamic consistency. Specifically, we keep the values of $P_L$ and $\mu_L$  unchanged while increasing the matching density as
\begin{equation}
n_L^{\rm new} = n_L + \delta n_B
\end{equation}
This approach naturally incorporates the definition of 1st-orderPTs: pressure and baryon chemical potential are continuous, while allowing discontinuous in baryon number density and energy density.

To quantify constraints on transition strength, we increase $\delta n_B$ from 0 to $1.5n_0$. The blue curve represents the case where $\delta n_B = n_0$ ( $n_L^{\rm new} = 2.5n_0$), while the green curve shows $\delta n_B = 1.5n_0$ ($n_L^{\rm new} = 3n_0$). Following the implementation of these 1st-order transitions, we recalculate the corresponding upper boundaries for each case. As $\delta n_B$ increases, the resulting upper boundary becomes progressively softer (pressure becomes smaller at a given energy density). This continues until reaching a critical threshold beyond which, for sufficiently large $\delta n_B$ around $2.5 n_0$, the matching conditions ($P_L$, $n_L^{\rm new}$, $\mu_L$) cannot connect with the quark EOS without violating either thermodynamic stability or causality constraints. This effectively establishes an upper limit on the allowable strength of 1st-orderPTs.

\begin{figure}[htp]
\centering
\includegraphics[width=1\hsize]{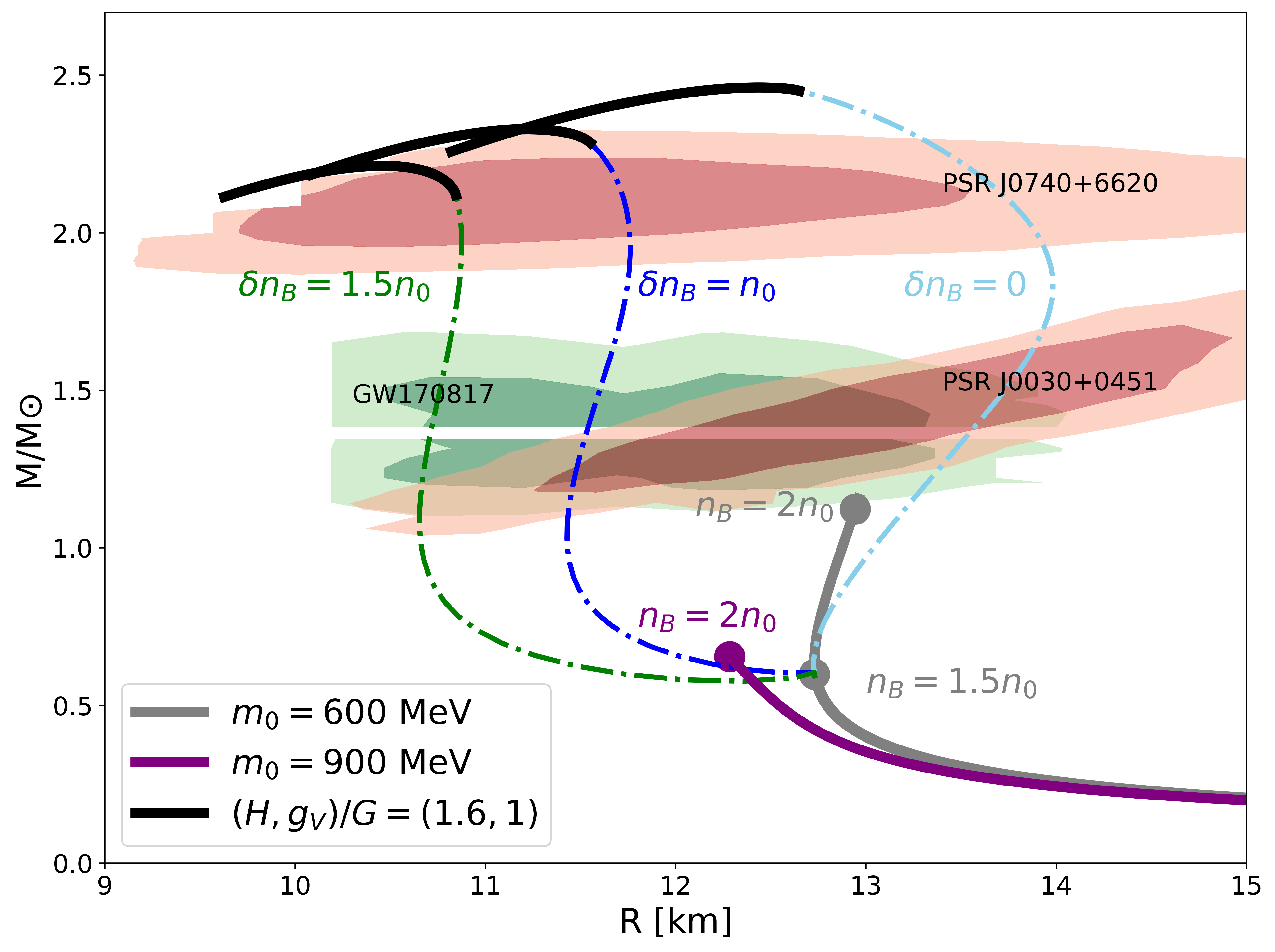}
\caption{$M$-$R$ relation for EOS connections between the hadronic phase described by the PDM with $m_0=600$ MeV (thick grey curve) and the quark phase with $(H, g_V)/G = (1.6, 1)$ (thick black curve). The colored curves represent the upper boundaries (stiffest possible EOS) for different strengths of 1st-orderPTs: $\delta n_B = 0$ (skyblue curve), $\delta n_B = n_0$ (blue curve), and $\delta n_B = 1.5n_0$ (green curve). }
\label{fig:m_r}
\end{figure}

\subsection{NS Properties and Observational Constraints}

By solving the Tolman-Oppenheimer-Volkoff (TOV) equations, we derive the $M$-$R$ relations corresponding to each upper boundary EOS with different $\delta n_B$ values, as shown in Fig.~\ref{fig:m_r}. These theoretical $M$-$R$ curves can be directly compared with recent NS observations, including NICER measurements of PSR J0030+0451 \citep{Miller:2019cac} ($M=1.44^{+0.07}_{-0.07} \,M_{\odot}$, $R = 13.7^{+2.6}_{-1.5}$ km) and PSR J0740+6620 \citep{Miller:2021qha} ($M=2.08^{+0.07}_{-0.07} \,M_{\odot}$, $R = 13.7^{+2.6}_{-1.5}$ km), as well as radius constraints derived from LIGO-Virgo gravitational wave observations \citep{LIGOScientific:2017vwq,LIGOScientific:2017ync,LIGOScientific:2018cki}.

Since we employ the upper boundary EOS representing the stiffest possible connection between hadronic phase and quark phase, if this maximally stiff EOS with a given $\delta n_B$ fails to satisfy observational constraints (e.g. cannot support the 2 $M_{\odot}$ mass; the radius is smaller than the observation constraints), we can conclusively rule out that transition strength. This provides a  method for constraining the allowable range of $\delta n_B$.

In Fig.~\ref{fig:m_r}, we present $M$-$R$ curves for pure hadronic EOSs with $m_0=600$ MeV (thick grey curve) and $m_0=900$ MeV (thick purple curve). As an example, we connect the hadronic EOS with $m_0=600$ MeV to the quark EOS with matching density $n_H=5n_0$ and $(H, g_V)/G = (1.6, 1)$ (thick black curve), showing results for transition strengths of $\delta n_B = 0$ (skyblue), $n_0$ (blue), and $1.5 n_0$ (green). As expected, increasing $\delta n_B$ leads to softer EOSs and smaller radii at a given mass. For $\delta n_B = 0$, while the corresponding $M$-$R$ curve has a radius at $1.4 M_{\odot}$ of 13.8 km, larger than the $1\sigma$ confidence region for GW 170817, we cannot exclude this transition strength as softer intermediate EOSs (between upper and lower bounds) could still satisfy all constraints. For  $\delta n_B = n_0$, all the observations are satisfied within $1\sigma$ level, indicating that PT strength $\delta n_B=n_0$ is still allowed. However, for $\delta n_B = 1.5n_0$, the $M$-$R$ curve violates the PSR J0030+0451 constraints at the $1\sigma$ level, and there do not exist even stiffer intermediate EOS. Therefore, requiring consistency with all observations at the $1\sigma$ confidence level excludes PTs with strength $\delta n_B \geq 1.5n_0$ for our selected hadronic and quark EOS parameters. 

\begin{figure*}
\centering
\begin{subfigure}[b]{0.48\textwidth}
    \centering
    \includegraphics[width=\textwidth]{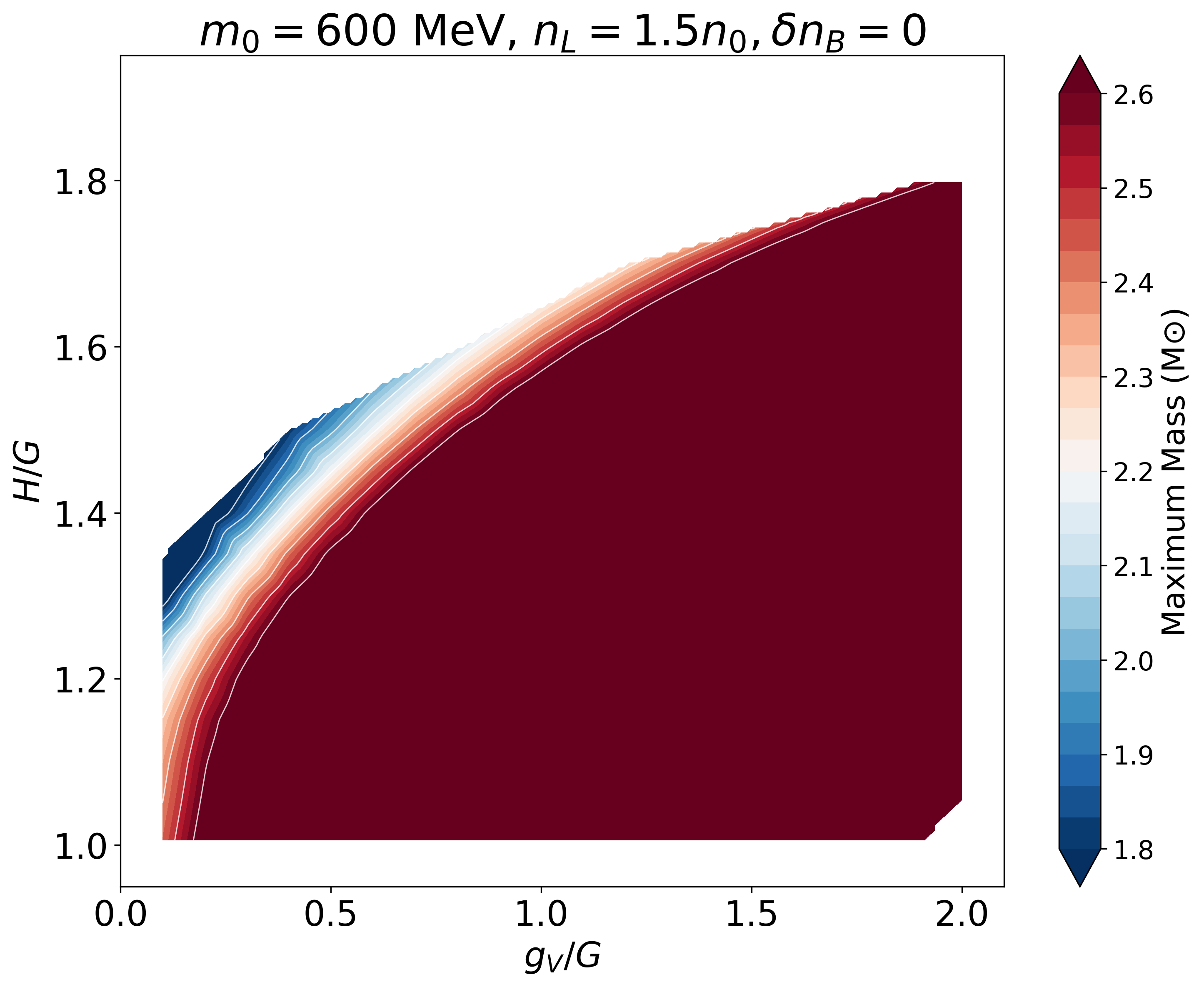}
    \caption{$\delta n_B=0$.}
\end{subfigure}
\hfill
\begin{subfigure}[b]{0.48\textwidth}
    \centering
    \includegraphics[width=\textwidth]{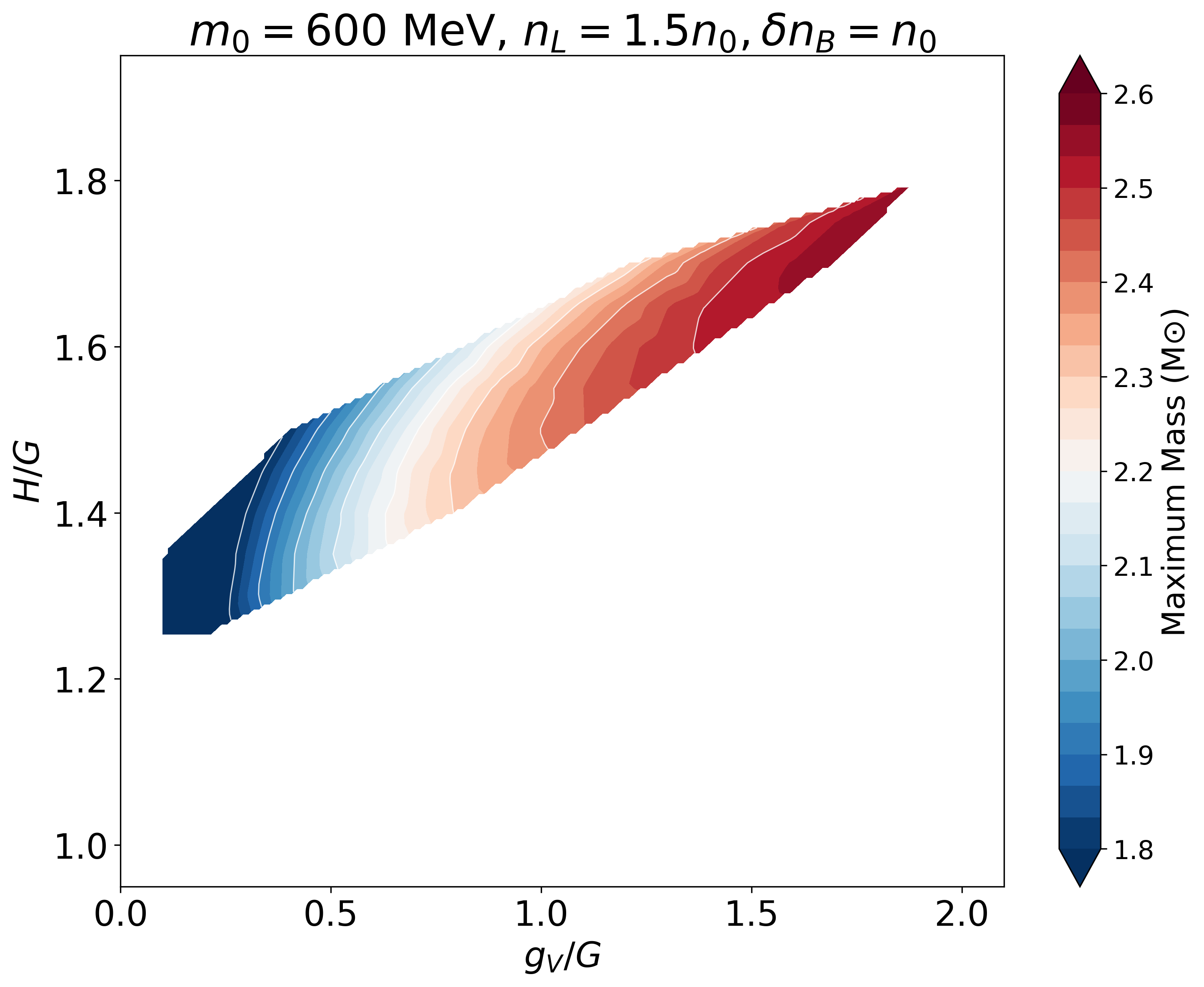}
    \caption{$\delta n_B=n_0$.}
\end{subfigure}

\begin{subfigure}[b]{0.48\textwidth}
    \centering
    \includegraphics[width=\textwidth]{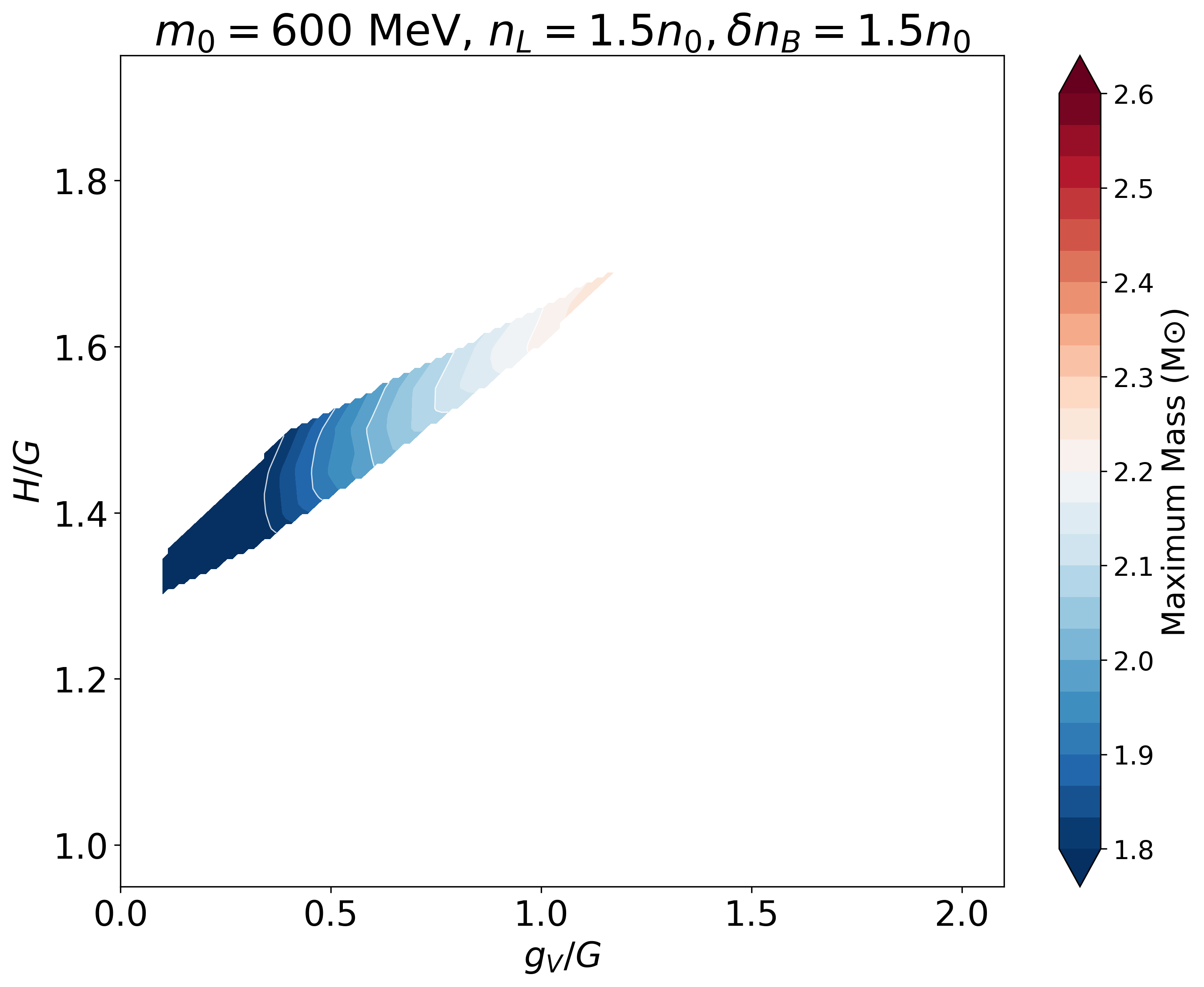}
    \caption{$\delta n_B=1.5 n_0$.}
\end{subfigure}
\hfill
\begin{subfigure}[b]{0.48\textwidth}
    \centering
    \includegraphics[width=\textwidth]{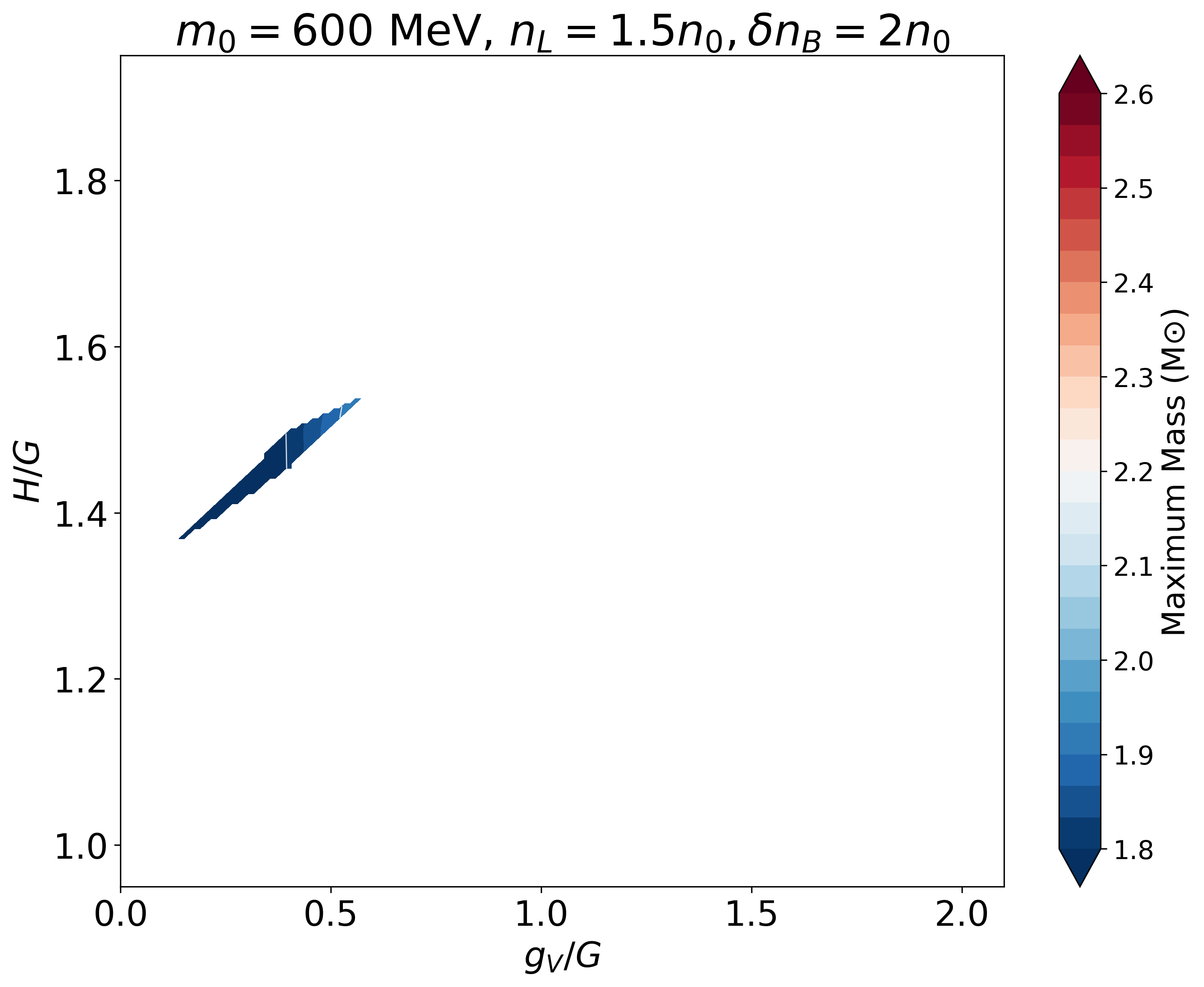}
    \caption{$\delta n_B=2n_0$.}
\end{subfigure}

\caption{Parameter space constraints for the NJL-type quark model with $m_0=600$ MeV and $n_L=1.5 n_0$. The color scale represents the maximum NS mass achievable with each $(H, g_V)/G$ parameter combination. Other parts without color indicate parameter combinations cannot connect with low density part.}
\label{fig:parameter}
\end{figure*}

While our previous analysis in Fig.~\ref{fig:p_e} and Fig.~\ref{fig:m_r} fixed the quark EOS parameters $(H, g_V)$  for demonstration purposes, we now systematically explore the full parameter space. For a given hadronic EOS with fixed $m_0$ and matching density $n_L$, the quark phase is characterized by  $H$ and  $g_V$ parameters. We examine how the allowable parameter region changes with increasing transition strength $\delta n_B$, as illustrated in Fig.~\ref{fig:parameter}. In each panel of Fig.~\ref{fig:parameter}, cross markers indicate parameter combinations that violate thermodynamic stability requirements. The color scale represents the maximum NS mass achievable with each $(H, g_V)/G$ parameter combination. As $\delta n_B$ increases, the allowed parameter region shrinks, and the corresponding maximum mass values decrease. At $\delta n_B = 2n_0$, only a small region of parameter space remains viable, and  none of these remaining parameter combinations can support NSs with masses over $2M_\odot$. This excludes transitions of strength $\delta n_B = 2n_0$ for our selected hadronic EOS with $m_0 = 600$ MeV and $n_L = 1.5n_0$.

For other choices of $\delta n_B$,  we first exclude parameter combinations that cannot support NSs with masses $\geq 2M_\odot$. Second, we apply radius constraints as illustrated in Fig.~\ref{fig:m_r} to find whether there still exist parameter spaces in quark phase. Through this parameter space analysis, we determine the maximum allowable 1st-order transition strength $\delta n_B$ and corresponding energy density discontinuity $\delta \varepsilon$ for our selected hadronic EOS. To satisfy all observational constraints within $1\sigma$ confidence level for our selected hadronic EOS, we find $\delta n_B \leq 1.12n_0$ with a corresponding energy density jump $\delta \varepsilon \leq 182 \text{ MeV/fm}^3$. If we relax our criteria to require consistency with observations at the $2\sigma$ level, the constraints become $\delta n_B \leq 1.68n_0$ with $\delta \varepsilon \leq 273 \text{ MeV/fm}^3$.

To complete our analysis, we examine how the maximum allowable strength of 1st-order PTs varies with different matching densities $n_L$ in the PDM. However, the applicability of pure hadronic descriptions becomes questionable at $n_B \geq 2n_0$ due to several physical considerations. First, nuclear many-body forces become increasingly relevant at these densities\cite{Baldo:1997ag,Li:2019xxz,Drischler:2021kxf}, suggesting that quark degrees of freedom may emerge before  quark matter formation\cite{Baym:2017whm,Baym:2019iky,Kojo:2021wax}. Second, additional complexity arises from potential hyperon appearance\cite{Lonardoni:2014bwa,Gal:2016boi} and exotic phases such as quarkyonic matter\cite{McLerran:2007qj, Duarte:2021tsx,Kojo:2021ugu,Fujimoto:2023mzy,Gao:2024jlp,Fujimoto:2024doc} in the intermediate density regime $2n_0 \leq n_B \leq 5n_0$. Given these limitations, we restrict our PDM analysis to $n_L \leq 2n_0$. It is worth noting that while nucleons remain the fundamental degrees of freedom at $n_B \leq n_0$ with no expected quark-hadron phase transitions, phenomenological models suggest that such transitions may occur at densities  slightly above $n_0$\cite{Ivanytskyi:2022mlk,Blaschke:2022egm,Gartlein:2023vif}. These early phase transitions would significantly impact NS $M$-$R$ relations and other observational signatures, making investigation of low-density transitions particularly relevant for astrophysical constraints.

\begin{figure}[htp]
\centering
\includegraphics[width=1\hsize]{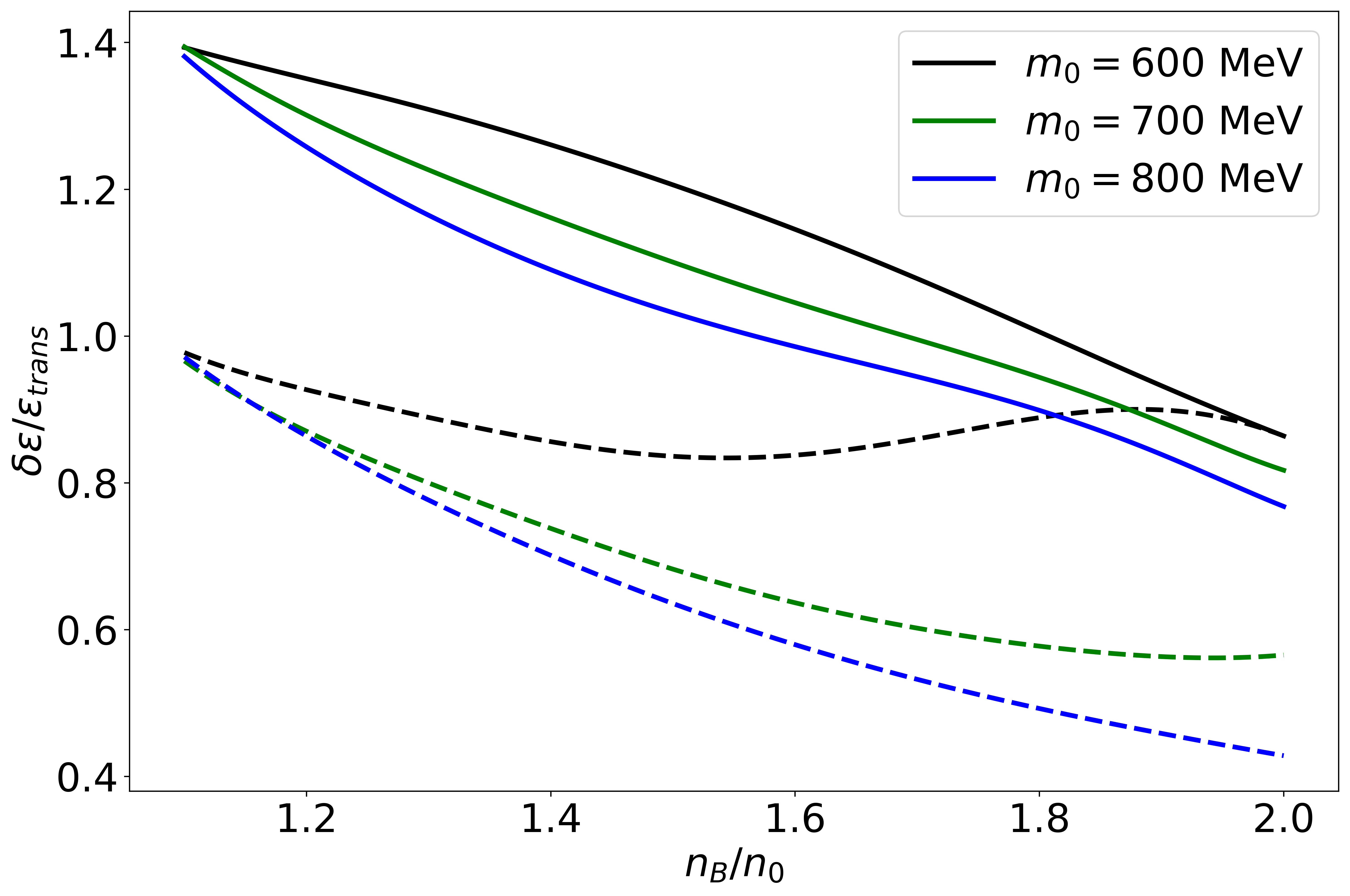}
\caption{Maximum allowable strength of 1st-orderPTs as a function of matching density $n_L$ with $n_H$ fixed to be $5n_0$.  Dashed curves represent constraints derived from NS  at $1\sigma$ confidence level, while solid curves show the less restrictive constraints at $2\sigma$ confidence level.  }
\label{fig:combine}
\end{figure}

To provide a physically meaningful measure of transition strength, we normalize the energy density jump by the energy density at the transition point since the absolute values of $\delta n_B$ and $\delta \varepsilon$ alone are difficult to interpret without a reference scale. Following Refs.~\cite{1983A, Alford:2015gna}, we therefore use the relative energy density jump $\delta \varepsilon/\varepsilon_{\rm trans}$, where $\varepsilon_{\rm trans}$ denotes the energy density at which the phase transition occurs. Fig.~\ref{fig:combine} presents the results of this systematic study across the range $n_L \leq 2n_0$ with $n_H=5n_0$. The dashed curves represent constraints derived from requiring consistency with NS observations at the $1\sigma$ confidence level, while solid curves indicate the less stringent $2\sigma$ confidence level constraints.n Fig.~\ref{fig:combine}, we find for a given density, the allowed PT strength decreases with increasing chiral invariant mass $m_0$. This behavior is expected since larger $m_0$ values lead to softer hadronic EOSs, making it more difficult to satisfy the $2M_{\odot}$ constraint when strong PTs are present. Also, at the lower density ($n_B \sim 1.1n_0$), the allowed PT strength becomes nearly degenerate across different $m_0$ values. This convergence occurs because the EOS at low densities is primarily determined by saturation properties, which are used as input parameters for all $m_0$ values, resulting in similar low-density behavior regardless of the chiral invariant mass. Furthermore, for stiff hadronic EOSs (e.g., $m_0 = 600$ MeV) at higher densities ($n_L \gtrsim 1.9n_0$), the constraints from $1\sigma$ and $2\sigma$ confidence levels converge. This degeneracy arises because the corresponding $M$-$R$ relations approach the observational limits from GW 170817 and PSR J0030+0451 (see Fig.~\ref{fig:m_r}), leaving less room for variation between different confidence levels.

\begin{figure}[htp]
\centering
\includegraphics[width=1\hsize]{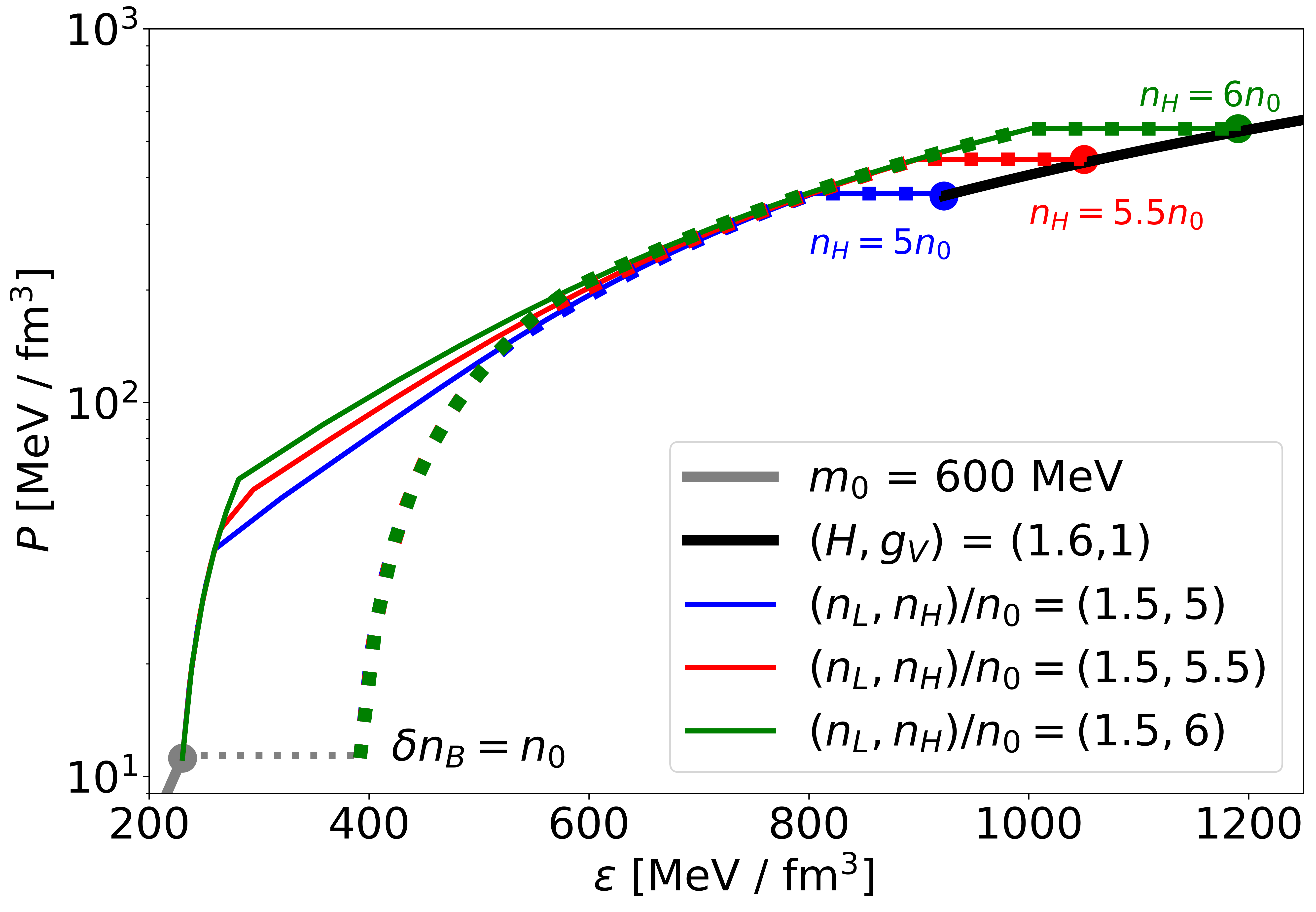}
\includegraphics[width=1\hsize]{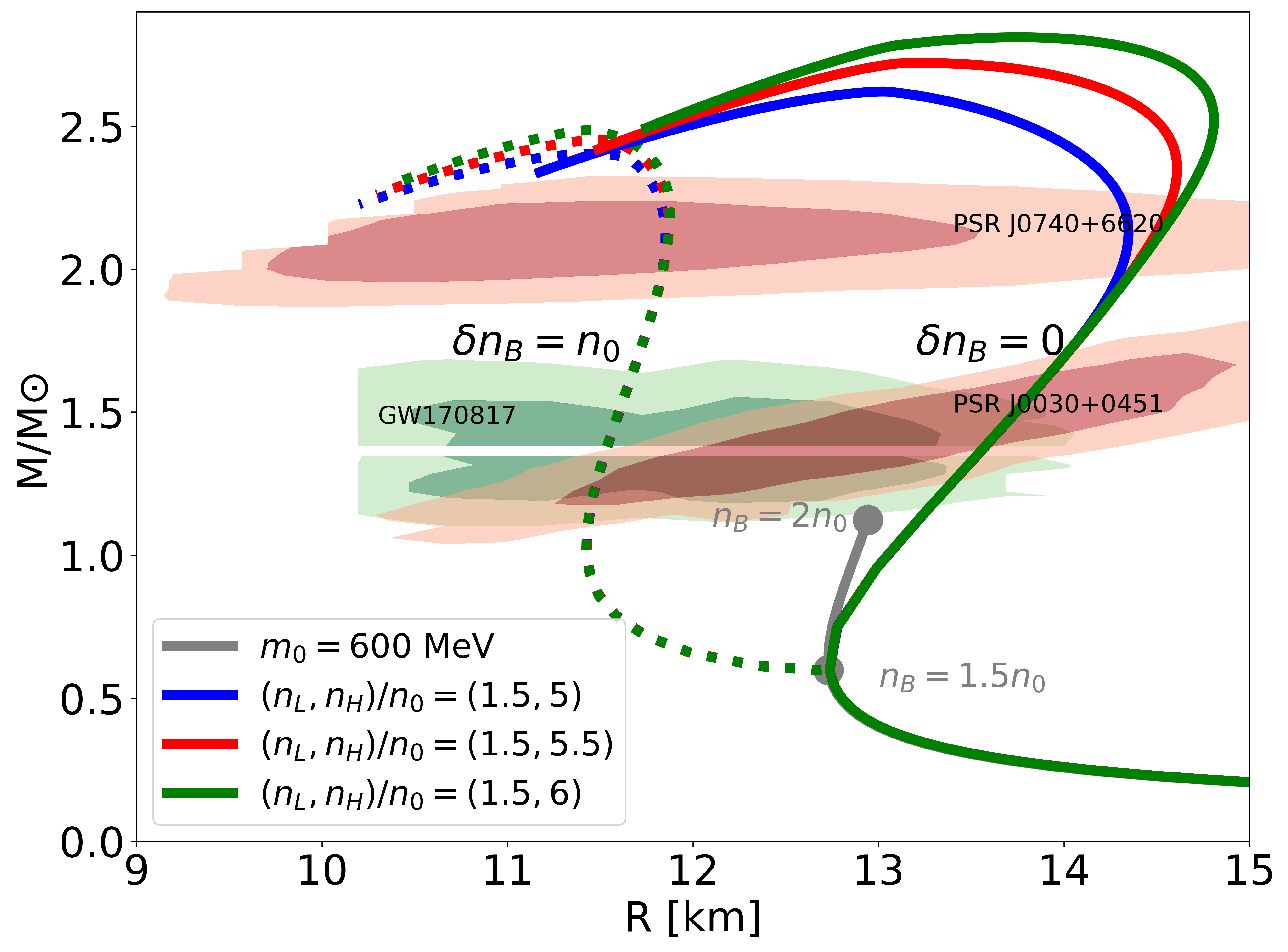}
\caption{Upper panel: EOS connections between the hadronic phase described by the PDM with $m_0=600$ MeV (grey curve) and the quark phase with $(H, g_V)/G = (1.6, 1)$ (black curve) in the pressure versus energy density plane for different high-density matching points. Lower panel: Corresponding mass-radius relations. Solid curves represent $\delta n_B = 0$ (no phase transition) and dashed curves represent $\delta n_B = n_0$ (1st-order phase transition). Color coding indicates different matching densities: blue for $(n_L, n_H)/n_0 = (1.5, 5)$, red for $(n_L, n_H)/n_0 = (1.5, 5.5)$, and green for $(n_L, n_H)/n_0 = (1.5, 6)$. }
\label{fig:am}
\end{figure}

Finally, we examine the sensitivity of our results to the choice of the high-density matching point by varying $n_H$. In the upper panel of Fig.~\ref{fig:am}, we show EOS connections between the hadronic phase described by the PDM with $m_0=600$ MeV (grey curve) and the quark phase with $(H, g_V)/G = (1.6, 1)$ (black curve) for different matching densities. The solid curves correspond to $\delta n_B = 0$ and the dashed curves to $\delta n_B = n_0$. We compare three different matching densities: $(n_L, n_H)/n_0 = (1.5, 5)$ (blue curves), $(n_L, n_H)/n_0 = (1.5, 5.5)$ (red curves), and $(n_L, n_H)/n_0 = (1.5, 6)$ (green curves). The results show that increasing $n_H$ has minimal impact on the low-pressure region of the EOS, where the curves remain identical (for $P < 40$ MeV in the solid curves and $P < 400$ MeV in the dashed curves). Beyond these pressure ranges, the EOSs become only slightly stiffer with larger $n_H$ values. The corresponding $M$-$R$ relations in the lower panel of Fig.~\ref{fig:am} demonstrate this insensitivity more clearly: for $\delta n_B = 0$ (solid curves), the radii are identical for masses below $1.7M_{\odot}$, while for $\delta n_B = n_0$ (dashed curves), the radii remain unchanged for masses below $2M_{\odot}$. Importantly, the radius at $1.4M_{\odot}$ is unaffected by changes in $n_H$ for both cases, validating our choice of $n_H = 5n_0$. While larger $n_H$ values lead to slightly higher maximum masses for the case $\delta n_B = n_0$, these changes do not significantly impact the main conclusions presented in Fig.~\ref{fig:combine}.

\section{SUMMARY AND DISCUSSION}\label{sec:summary}

In this work, we have  investigated the constraints on the strength of 1st-order phase transitions in NS matter, establishing a  framework that connects the microscopic origin of nucleon mass to macroscopic NS observables. To construct the EOS relevant to NS densities, we combine three theoretical frameworks to describe different density regimes in NS matter. For the hadronic phase at low densities, we employed the PDM, which provides a natural framework for understanding chiral symmetry restoration in dense matter. In this model, the mass of nucleon and its parity partner become degenerate when chiral symmetry is restored, with their mass splitting controlled by the chiral condensate. A key parameter in the PDM is the chiral invariant mass $m_0$, which represents the portion of the nucleon mass that persists even when chiral symmetry is fully restored. This parameter profoundly influences the EOS in the PDM: larger values of $m_0$ lead to weaker coupling to the $\sigma$ and $\omega$ mesons, resulting in a softer EOS at supranuclear densities. By varying the values of $m_0$, we systematically explored how different choices of the chiral invariant mass affect the nuclear EOS and, consequently, the allowed properties of 1st-order phase transitions.

At high densities relevant to neutron star cores, where the baryon chemical potential reaches $\mu_B \sim 1.5-2$ GeV and the system enters a strongly coupled regime beyond perturbative QCD, we utilized the Nambu-Jona-Lasinio model to describe quark matter. This effective theory captures essential QCD features including dynamical chiral symmetry breaking and color superconductivity through four-fermion interactions, characterized by the vector coupling $g_V$ controlling repulsive interactions and the diquark coupling $H$ governing attractive correlations that lead to color superconductivity.

We use the PDM up to the matching density $n_L$ and employ the NJL-type quark model from the high-density point $n_H$ onwards, and crucially, for the intermediate density region between $n_L$ and $n_H$, we utilize the constrained upper boundary within the integral constraint framework as introduced in Ref.~\cite{Komoltsev:2021jzg}. This model-independent approach requires only fundamental principles of thermodynamic stability and causality to determine physically allowable connections between the low and high-density phases. The framework  defines upper and lower boundaries in the pressure-energy density plane, representing the stiffest and softest possible connections, respectively, making it ideally suited for constraining the largely unknown density regime between $2n_0$ and $5n_0$ where the quark-hadron transition likely occurs.

Within this framework, we model first-order phase transitions by introducing a density discontinuity while maintaining thermodynamic consistency. Specifically, we keep the pressure $P_L$ and chemical potential $\mu_L$, determined from the PDM, at the transition point fixed while increasing the matching density from $n_L$ to $n_L + \delta n_B$. This approach naturally incorporates the defining characteristics of first-order transitions: continuous pressure and chemical potential with discontinuous baryon number density and energy density. The strength of the phase transition is then characterized by the magnitude of the density jump $\delta n_B$ or, more meaningfully, by the relative energy density jump $\delta\varepsilon/\varepsilon_{\text{trans}}$.

By systematically increasing the transition strength and comparing the resulting $M$-$R$ relations with NS observations, we established constraints on the maximum allowable $\delta\varepsilon/\varepsilon_{\text{trans}}$.  These constraints depend on the choices of chiral invariant mass: larger values of $m_0$ lead to softer hadronic EOSs, which in turn impose tighter restrictions on the phase transition strength. This correlation directly links the fundamental question of the origin of nucleon mass to observable neutron star properties, providing a unique astrophysical probe of chiral dynamics.

To assess the robustness of our results, we also examined the sensitivity to the choice of matching points $n_H$ in the NJL-type quark model . Varying the high-density matching point $n_H$ from $5n_0$ to $6n_0$ produces only minor changes in the $M$-$R$ relations, with the radius at $1.4M_\odot$ remaining essentially unchanged and maximum masses showing only slight increases. This stability against reasonable variations in matching points strengthens confidence in our conclusions.

Looking toward future developments, several extensions of this work merit investigation. Since we restricted our analysis to $n_L \leq 2n_0$ due to concerns about the reliability of purely hadronic descriptions at higher densities, future work could extend the PDM framework to include hyperon effects, allowing investigations at higher densities while maintaining theoretical consistency. Additionally, the framework could be expanded to include exotic phases such as quarkyonic matter, which may provide important intermediate states in the transition from hadronic to quark matter.

\begin{acknowledgments}
The author gratefully acknowledges Masayasu Harada and Atsushi Hosaka for valuable correspondence and discussion.
\end{acknowledgments}

\newpage

\appendix \label{app:quark}
In this appendix, we construct neutron star matter EOS in quark matter part.

\section{Quark matter EOS} 

Following Refs.~\cite{Baym:2017whm,Baym:2019iky}, we use the NJL quark model to describe the quark matter. 
The model includes three-flavors and U(1)$_A$ anomaly effects through the quark version of the KMT interaction. 
The coupling constants are chosen to be the Hatsuda-Kunihiro parameters 
which successfully reproduce the hadron phenomenology at low energy (\cite{Baym:2017whm, Hatsuda:1994pi}): 
$G\Lambda^{2}=1.835, K\Lambda^{5}=9.29$ with $\Lambda=631.4\, \rm{MeV}$, see the definition below.
The couplings $g_{V}$ and $H$ characterize the strength of the vector repulsion and attractive diquark correlations whose range will be examined later 
when we discuss the NS constraints.

We can then write down the thermodynamic potential as
\begin{equation}
\begin{aligned}
\Omega_{\mathrm{CSC}}
=&\, \Omega_{s}-\Omega_{s}\left[\sigma_{f}=\sigma_{f}^{0}, d_{j}=0, \mu_{q}=0\right] \\
&+\Omega_{c}-\Omega_{c}\left[\sigma_{f}=\sigma_{f}^{0}, d_{j}=0\right],
\end{aligned}
\end{equation}
where 
the subscript 0 is attached for the vacuum values, and
\begin{equation}
\begin{aligned}
&\Omega_{s}=-2 \sum_{i=1}^{18} \int^{\Lambda} \frac{d^{3} \mathbf{p}}{(2 \pi)^{3}} \frac{\epsilon_{i}}{2} \label{energy eigenvalue},\\
&\Omega_{c}=\sum_{i}\left(2 G \sigma_{i}^{2}+H d_{i}^{2}\right)-4 K \sigma_{u} \sigma_{d} \sigma_{s}-g_{V} n_{q}^{2},
\end{aligned}
\end{equation}
with $\sigma_{f}$ are the chiral condensates, $d_{j}$ are diquark condensates, and $n_{q}$ is the quark density. 
In Eq.(\ref{energy eigenvalue}), $\epsilon_{i}$ are energy eigenvalues obtained from inverse propagator in Nambu-Gorkov bases
\begin{equation}
S^{-1}(k)=\left(\begin{array}{lc}
\gamma_{\mu} k^{\mu}-\hat{M}+\gamma^{0} \hat{\mu} & \gamma_{5} \sum_{i} \Delta_{i} R_{i} \\
-\gamma_{5} \sum_{i} \Delta_{i}^{*} R_{i} & \gamma_{\mu} k^{\mu}-\hat{M}-\gamma^{0} \hat{\mu}
\end{array}\right),
\end{equation}
where
\begin{equation}
\begin{aligned}
&M_{i} =m_{i}-4 G \sigma_{i}+K\left|\epsilon_{i j k}\right| \sigma_{j} \sigma_{k}, \\
&\Delta_{i} =-2 H d_{i} ,\\
&\hat{\mu} =\mu_{q}-2 g_{V} n_{q}+\mu_{3} \lambda_{3}+\mu_{8} \lambda_{8}+\mu_{Q} Q,\\
&(R_{1}, R_{2}, R_{3})=(\tau_{7}\lambda_{7}, \tau_{5}\lambda_{5}, \tau_{2}\lambda_{2}).
\end{aligned}
\end{equation}
$S^{-1}(k)$ is $72\times72$ matrix in terms of the color,
flavor, spin, and Nambu-Gorkov basis, which has 72 eigenvalues. $M_{u,d,s}$ are the constituent masses of $u, d, s$ quarks and $\Delta_{1,2,3}$ are the gap energies. 
The $\mu_{3,8}$ are the color chemical potentials which will be tuned to achieve the color neutrality. 
The total thermodynamic potential including the effect of leptons is 
\begin{equation}
\Omega_{\mathrm{Q}}=\Omega_{\mathrm{CSC}}+\sum_{l=e, \mu} \Omega_{l}.
\end{equation}
The mean fields are determined from the gap equations,
\begin{equation}
0=\frac{\partial \Omega_{\mathrm{Q}}}{\partial \sigma_{i}}=\frac{\partial \Omega_{\mathrm{Q}}}{\partial d_{i}},
\end{equation}
From the conditions for electromagnetic charge neutrality and color charge neutrality, we have
\begin{equation}
n_{j}=-\frac{\partial \Omega_{\mathrm{Q}}}{\partial \mu_{j}}=0,
\end{equation}
where $j = 3,8, Q$. 
The baryon number density $n_{B}$ is determined as
\begin{equation}
n_{q}=-\frac{\partial \Omega_{\mathrm{Q}}}{\partial \mu_{q}},
\end{equation}
where $\mu_{q}$ is $1/3$ of the baryon number chemical potential. After determined all the values, we obtain the pressure as
\begin{equation}
P_{\mathrm{Q}}=-\Omega_{\mathrm{Q}}.
\end{equation}
%
\bibliography{sample631}

\end{document}